\documentclass[12pt]{article}
\usepackage{graphicx}
\usepackage{epsf}

\def\f{\frac}
\def\b{\beta}

\def\n{\nu}
\def\m{\mu}

\def\t{\theta}

\def\be{\begin{equation}}
\def\ee{\end{equation}}
\def\bea{\begin{eqnarray}}
\def\eea{\end{eqnarray}}
\def\a{\alpha}
\def\ba{\begin{array}}
\def\ea{\end{array}}

\def\l{\left}
\def\r{\right}

\begin{document}
\begin{titlepage}
\begin{center}
{\Large \bf William I. Fine Theoretical Physics Institute \\
University of Minnesota \\}
\end{center}
\vspace{0.2in}
\begin{flushright}
FTPI-MINN-09/39 \\
UMN-TH-2822/09 \\
October 2009 \\
\end{flushright}
\vspace{0.3in}
\begin{center}
{\Large \bf Photon-stimulated production of electron-positron pairs in electric field  \\}
\vspace{0.2in}
{\bf A. Monin \\}
School of Physics and Astronomy, University of Minnesota, \\ Minneapolis, MN
55455, USA, \\
and \\
{\bf M.B. Voloshin  \\ }
William I. Fine Theoretical Physics Institute, University of
Minnesota,\\ Minneapolis, MN 55455, USA \\
and \\
Institute of Theoretical and Experimental Physics, Moscow, 117218, Russia
\\[0.2in]
\end{center}

\begin{abstract}
The rate of $e^+e^-$ pair creation by external electric field in the presence of an incident photon beam is calculated for the photon energy far below the threshold, $\omega \ll m$, and the field strength small as compared to the critical one, $e E \ll m^2$. We find the pair production rate using a recently developed method based on calculation of the process in a thermal bath with subsequent identification of the contribution of single-photon states. We demonstrate that a non-trivial dependence on the ratio of the small parameters, $\omega m/(e E)$, emerges in this approach from an essentially (semi)classical calculation. 
\end{abstract}

\end{titlepage}
\newpage

\section{Introduction.}
It has been theoretically understood since long ago~\cite{Sauter} that a static electric field $E$ can spontaneously produce electron-positron pairs due to quantum tunneling. The probability of this phenomenon, usually called the Schwinger process, can be found from the imaginary part of the one-loop effective action in an external field~\cite{Heisenberg:1935qt, Schwinger}, and the pair production rate per unit volume in a constant field is given by the well known formula
\be
\Gamma = \f {( e E ) ^ 2} {4 \pi ^3} \sum _ {n = 1} ^ { + \infty } \f {1} {n ^ 2} \exp \l ( - \f {\pi \, m ^ 2 } {e E} \, n \r )~,
\label{ehs}
\ee
with $e$ and $m$ being respectively the charge and the mass of the electron.
The generalizations of this formula to varying external field include the special cases of a spatially constant field with the time dependence $E \sim 1/ \cosh ^ 2 \l ( \Omega t \r )$~\cite{Narozhnyi:1970uv} and of the field  pointing along the $z$ axis and arbitrarily depending on the light cone variable $ E _z ( t  \pm x  ) $~\cite{Fried:2001ur}, a complete review of the topic can be found in Ref.~\cite{Dunne:2004nc, Dittrich:2000zu}.

Being thoroughly investigated theoretically, the Schwinger process has no experimental evidence so far. The reason is that any practically available strength  of the electric field is much smaller than the critical value $E_c=m^2/e \sim 10^{16}\,$V/cm at which the probability described by Eq.(\ref{ehs}) would not be exponentially suppressed. 

It has been suggested recently~\cite{sgd} that the pair creation can be significantly stimulated by superimposing a relatively weak photon beam with a (quasi)static electric field. It was shown~\cite{Dunne:2009gi} that in the presence of an external photon the barrier for the tunneling is effectively lowered and the negative exponential power in the pair production rate is decreased in absolute value. In particular, for a photon with energy $\omega$ propagating perpendicularly to the field $E$ 
the negative exponential power in the pair production rate at the threshold $\omega=2m$ is modified
from $- \pi \, m^2/(eE)$ to $- (\pi-2)\, m^2/(eE)$, leading to a large exponential
enhancement of the rate.

In the present paper we consider the photon induced pair creation in an external electric field in the realistic limit $E \ll E_c$ and at lower photon energies $\omega \ll m$ for which higher beam intensity can be practical. Under this condition the leading effects are described by the so-called Keldysh parameter $\gamma_\theta = m \, \omega \, \sin \theta/(eE)$, with $\theta$ being the angle between the photon momentum and the electric field $E$. We do not assume the Keldysh parameter to be small and find the exact in this parameter expression for the attenuation rate $\kappa$ for the photon beam intensity due to the pair production in the form
\be
 { \kappa }_\parallel ( \vec k )  =  2\, \f {\a \, m^2} {\omega} e ^  { - \f {\pi \, m ^ 2 } {e E} }  
\l [ I _1 \l ( \gamma _ \t  \r ) \r ] ^ 2~,
\label{kba}
\ee
with $I_1(x)$ being the standard notation for the modified Bessel function. 
Our consideration here is restricted to the lowest order in the ratio $\omega \, \sin \theta/m$, and in this order we find that only the photons whose polarization is parallel to $\vec E$ stimulate the pair production (hence the notation ${ \kappa }_{\parallel}$), while the effect for the orthogonal polarization, $\kappa_\perp$, arises only in a higher order in this ratio. The exponential behavior of the Bessel function in Eq.(\ref{kba}) at large argument matches the low $\omega$ limit of the exponential expression found in Ref.~\cite{Dunne:2009gi}, so that our result describes both the exponential and the pre-exponential factors in this limit.

The photon-induced pair creation can be calculated in a standard way in terms of the imaginary part of the electromagnetic vacuum polarization function $\Pi$ in external field:
\be
\kappa = - {1 \over \omega} \, \Im \Pi~
\label{kapi}
\ee 
starting from the known~\cite{Dittrich:2000zu} general formulas for $\Pi$. In fact the expression (\ref{kba}) is very recently found~\cite{Dunne:2009gi} in this way by considering a small $\omega$ expansion of the contour integral representation of the imaginary part of the polarization tensor for
photons in a constant E field.  
In this paper we do not use this approach but rather find the result (\ref{kba}) by considering the Schwinger pair creation as a semiclassical tunneling process\footnote{Which treatment in fact goes back to the original idea~\cite{Sauter}}, and using the Euclidean-space description of the tunneling trajectory~\cite{Affleck, Dunne:2005sx, Dunne:2006st}. In such semiclassical approach, instead of summing the loop graphs for the vacuum-to-vacuum transition amplitude, one uses the so-called bounce~\cite{Coleman} trajectory in the Euclidean space time similarly to the methods used in treatment of false vacuum decay~\cite{vko,Coleman}. In order to find  the pair creation rate stimulated by photons we use the recently developed~\cite{mvsp,mvsw} extension of this approach using a thermal calculation of the tunneling rate. An appropriate interpretation of the result for the probability of the process at finite temperature in a thermal bath allows to extract the behavior of the rate induced by individual particles present in the bath.
(In fact the technique allows to find the probability of pair production induced by arbitrary number of particles).

In what follows we provide the actual calculation leading to the expression (\ref{kba}). In the Section 2 we briefly recapitulate the quasiclassical method of calculating the probability rate, and in the Section 3 derive the expression for the rate at nonzero temperature $T$ in terms of expansion in powers of $m T/(eE)$. In the Section 4 we relate the thermal result to the contribution of the one-photon induced process and thus we find the probability described by Eq.(\ref{kba}). Finally, the Section 4 contains the discussion and concluding remarks.

\section{Euclidean-space tunneling}
The Euclidean space approach~\cite{Coleman} to tunneling is based on constructing a localized solution to the classical equations of motion, which solution is called a bounce, and the exponential factor in the rate is determined by the Euclidean action on the bounce $S_{B}$ as $\Gamma \propto \exp(-S_B)$, while the pre-exponential factor is derived~\cite{cc} from a calculation of the Euclidean path integral around the bounce trajectory. It can be also mentioned that if in the problem there is a separation of scales such that some degrees of freedom can be considered as soft on the scale of the size of the bounce, both the exponential factor~\cite{Coleman} and the pre-exponential one~\cite{Voloshin:1985id} can be treated within an effective theory of those soft variables.

In order for an electron-positron pair to be created, the electric field $E$ has to produce the work equal to $2m$. This requires the length $\ell = 2m/(eE)$, and at a field weaker than the critical, $E \ll m^2/(eE)$ the length scale $\ell$ greatly exceeds the electron Compton wave length, $\ell \gg m^{-1}$. In this situation the tunneling bounce configuration in the problem of pair creation can be treated within an effective low-energy theory with essentially classical action for the electrons in external electromagnetic field:
\be
S = m \int dl - e \int A _ \m \, d x ^ \m,
\label{effa}
\ee
with $x _ \m$ being the coordinate of the particle and $dl$ is the element of the length of the particle trajectory. In a constant electric field in the $x$ direction, $E=E_x$, one can write the potential as $A_0=A_{ext}=E \, x$ and thus rewrite the action (\ref{effa}) for a closed trajectory in terms of its length $L$ and the area $A$ that it encircles:
\be
S = m L - eE A.
\ee
Thus the trajectory extremizing the action, the bounce, is a circle in the $(t,x) $ plane with the radius
\be
R = \f {m} {eE}
\label{radius}
\ee
as shown in Fig.~1.
(The diameter of the circle is such that the work produced by the constant electric field on that distance is equal to the total mass of the pair produced $2 m$.)  
\begin{figure}[ht]
  \begin{center}
    \leavevmode
    \epsfxsize=3cm
    \epsfbox{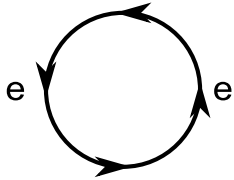}
    \caption{The bounce configuration for $e^+e^-$ pair creation.}
  \end{center}
\label{inst}
\end{figure}
 The value of the action on this trajectory is  $S _B = \f {\pi \, m^2} {eE}$, in a complete agreement with the leading exponent in the exact expression (\ref{ehs}) under the condition $E \ll E_c$, which ensures applicability of the semiclassical treatment.

\section{Pair creation in electric field in a thermal bath.}
As mentioned, we eventually find the photon-induced pair production rate by extracting the corresponding one-particle contribution from an expression for the Schwinger process at a finite temperature. Therefore we start with calculating the probability rate per unit volume at nonzero temperature. For a sufficiently small temperature, namely $T \ll m$, one can still employ the same effective Euclidean one particle action (\ref{effa}), except that now the system lives on a cylinder with a periodic Euclidean time with the period equal to the inverse temperature $\b = 1/T$. Equivalently one can consider the system on the (t,x) plane with periodic in $t$ boundary conditions (see Fig.2).
\begin{figure}[ht]
  \begin{center}
    \leavevmode
    \epsfxsize=10cm
    \epsfbox{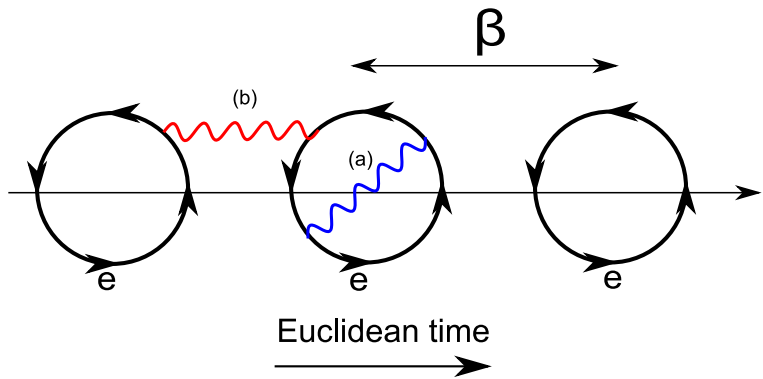}
    \caption{Periodic plane. Two types of correction: (a) is the correction due to the self interaction of the particle  on a circle; (b) is the correction due to the interaction of the circles separated in Euclidean time.}
  \end{center}
\label{poten}
\end{figure}
Moreover for the purpose of the present calculation it is sufficient to consider arbitrarily low but non vanishing temperature, and we thus also impose the condition $T < (2R) ^ {-1} ={eE}/({2m})$ which even more justifies the applicability  of the effective action and also ensures that the circular bounce fully fits within one period. 

The action for the bounce and thus the probability of pair creation does not change in this limit if the electromagnetic field is considered as an external object without dynamics of its own. We however are interested in the effects produced by the photons in the thermal bath and we should thus consider the dynamics of the electromagnetic field by adding to the low energy action the kinetic term for the field $A_\m$. In order to exclude the contribution of the energy of the external field to the action one can make the shift in the definition of the electromagnetic potential: $A_\m \to A_{ext}+A_\m$ and write the effective action as
\be
S[x,A] = m \int dl - e E A - e \int A _ \m d x ^ \m - {1 \over 4} \, \int d ^ 4 x \, F _ {\m \n} ^ 2~,
\label{effap}
\ee
where $A_\m$ is the shifted potential with the corresponding field strength $F_{\m \n}=\partial_\m A_\n - \partial_\n A_\m$. 

Generally, the rate is found~\cite{Dunne:2004nc, Affleck, Coleman, Voloshin:1985id} by calculating the partition function around the one bounce configuration
\be
\Gamma = \f { 2 } {VT} \mathrm{Im} \int \mathcal{D} x _ \m \, \mathcal{D} A _ \m \exp  {- S [x,A]} .
\ee
with effective one particle Euclidean action $S[x,A]$ given by the expression (\ref{effap}), and $VT$ being the space-time volume of the system.

The zero temperature result corresponds to the limit $\b \to \infty$, and, without any corrections from exchange of the photons, yields  the well known expression for the rate
\be
\f {\Gamma} {V} = \f {( e E ) ^ 2} {4 \pi ^3} \, \exp \l ( - \f {\pi \, m ^ 2 } {e E} \r ),
\label{gam1}
\ee
which corresponds to the first term in the sum (\ref{ehs}).
The self interaction of the particle (one-loop correction: the correction of the type (a) in Fig.~2) was taken into account in paper \cite{Affleck}. It amounts to a finite additive term in the exponent, $ e ^ 2 / 4 = \pi \a$. The thermal effect that eventually leads to the result in the present paper can in fact be viewed as a thermal distortion of this self-interaction term due to the modification of the photon propagator on the cylinder as compared to an infinite space-time. In the equivalent periodic picture of Fig.~2 this modification can be considered as an interaction between the periodic copies of the current loops with the photon propagator being that in an infinite space-time (the corrections of type (b)).
 
Before proceeding to a calculation of this latter effect we note that the corrections due to the thermal fluctuations of the shape of the bounce, i.e. deviations from the circle, are of a higher order in $ {eE}/ { m ^2}$ and are entirely neglected in the present treatment.

The contribution to the action due to such interaction has the following form
\be
\Delta S _ {tot} = - \int d ^ 4 \, x \l ( \f {1} {4} F _ {\mu \nu} ^ 2 + A _ \mu \, j _\mu \r ) = - \f {1} {2}  \int d ^ 4 \, x
A _ \mu \, j _\mu.
\ee
with $j _ \mu$ being the total current in the circles
\be
j _ \mu = \sum _ n  e \, n  ^ { ( n ) } _ \m \delta ( r _ n - R ) \, \delta(y) \, \delta(z),
\ee
and $n ^ {(n)} _ \m $ is the tangential unit vector to the $n-$th circle:
\be
n ^ {(n)} _ \m = ( - \sin \t _ n, \cos \t _ n)~,
\ee
where $(r_n, \theta_n)$ are the polar coordinates with the origin at 
the center of the $n$-th copy of the circle located at $(x,t) = (0, n \, \beta)$. $A  _ \m$ is the field produced by all those circles
\be
A  _ \m = \sum _ {n} A ^ { (n)} _ \m,
\ee
where all $A _ \m ^ { (n) }$ in turn are the the solutions of the Laplace equation
\be
\Delta A ^ {(n)} _ \m = j ^ {(n)} _ \m, 
\ee
\be
A ^ {(n)}_ \m (r _ n , \t _ n, y=0, z=0) = \f {e R^2} {2\pi} \, \f {n ^ {(n)} _ \m} {r _ n \, ( r _ n ^ 2 - R ^ 2 ) },
\label{pota}
\ee

Omitting the contribution from $A ^ { (0)} _ \m$, which corresponds to self interaction, we get the correction to the action per one period $\b$
\bea
\Delta S & = & -  \sum _ {n = 1} ^ { + \infty} \int d ^ 4 \, x \, A ^ {(n)}_ \m \, j ^ {(0)} _ \m = -  {e ^ 2 \over 2} \, \sum _ {n = 1} ^ { + \infty}  \l ( \f {1 - 2 (RT/n) ^ 2 } { \sqrt { 1 - (2 \, RT/n) ^ 2} } - 1 \r ) \nonumber \\
& = & - { e ^ 2 \over 2} \, \sum _ {p = 2} ^ {\infty}
2 ^ {2 p - 1} (RT) ^ {2 p}\, \f { (p-1) \, \Gamma (p - 1/2) } { \sqrt { \pi } \, \Gamma (p+1)} \, \zeta (2 p) ~,
\label{dels}
\eea
where the sum runs over only positive $n$ and it is taken into account that the contribution of the terms with negative $n$ is the same as that from $n>0$. Finally, $\zeta$ is the standard Riemann zeta function
$$\zeta(q)=\sum_{n=1}^\infty n^{-q}~.$$ 

A remark is in order concerning the apparent `extra' factor of one half in Eq.(\ref{dels}). In the treatment in `flat' space time with periodic copies this factor arises for the following reason. Each term $-A ^ {(n)}_ \m \, j ^ {(0)} _ \m$ in the sum corresponds to the additional action within {\em the pair} of the $n-th$ and $0-th$ current loops. Thus the additional action per one loop, i.e. per one period $\b$, is one half of that. In the picture of a current loop on the cylinder the equivalent explanation of this factor is that the self-interaction of the loop through $n$ windings of the photon propagator around the cylinder does not contain any notion of the sign of $n$. Therefore summing over the positive and negative values of $n$ would be double counting.

Using the pre-exponential factor from Eq.(\ref{gam1}) and the expression (\ref{dels}) for $\Delta S$, one can write the  rate of pair creation at finite temperature in the form
\be
\f { d \Gamma _ T} {d V}  =  \f {( e E ) ^ 2} {4 \pi ^3} \, \exp \l ( - \f {\pi \, m ^ 2 } {e E} - \Delta S\r )~,
\label{temp_rate}
\ee
so that the thermal enhancement factor is $\exp(-\Delta S)$.

\section{Pair production induced by a photon.}

In a microscopic description of the thermal effects, the enhancement of pair creation in a bath at finite temperature arises through the stimulation of the process by the photons present in the bath. The dependence of the photon induced process on the photon energy $\omega$ then translates into the dependence on the temperature $T$ after averaging over the thermal distribution of the photons with the standard density function 
\be
n ( \vec{k} ) = \f {1} {e ^ {\omega  \b} - 1 }
\label{photdis}
\ee 
with $\omega = |\vec k|$. The number of photons involved in each of these microscopic processes can be readily identified by the power of the factor $e^2$. Since the thermal correction (\ref{dels}) in the action is proportional to $e^2$, the one-photon contribution to the thermal rate is given by the linear in $\Delta S$ term in the expansion of the factor $\exp(-\Delta S)$ in the expression (\ref{temp_rate}). Namely, the one-photon contribution to the pair creation rate in a thermal state is given by~\footnote{In Refs.~\cite{mvsp, mvsw} the contribution of processes with different number of massless bosons in a thermal bath was separated by formally introducing a negative chemical potential. In the problem discussed here this is not necessary, since the power of the coupling $e^2$ automatically `tags' the number of photons.}
\be
{\Gamma _ {1 \gamma} / V} = \f {( e E ) ^ 2} {4 \pi ^3} \, \exp \l ( - \f {\pi \, m ^ 2 } {e E} \r ) \, (- \Delta S)~.
\label{g1_e}
\ee

On the other hand the same contribution can be found in terms of  the averaged over the photon polarizations probability $\bar \kappa$ rate of pair production induced by a photon, which is the same as the absorption rate for the photons per unit length. The latter can be expanded in a power series of $\omega \sin \t$, with yet to be defined coefficients $C _n$ as~\cite{Dunne:2009gi}
\be
\bar { \kappa } ( \vec {k} ) = \f {1} {\omega} \sum _ {p = 2} ^ {\infty} C _ { p } (\omega \, \sin \t) ^ {2 p - 2}. 
\ee
The functional form of $\kappa$ in fact follows from its relation (\ref{kapi}) to the vacuum polarization in electric field and the dependence of the on-shell imaginary part $\Im \Pi$ on $\omega \sin \theta$~\cite{Dittrich:2000zu}.

The thermal probability is then found in terms of the coefficients $C _ p$ by integrating over the photon momentum with the weight given by the distribution (\ref{photdis}):
\be
{\Gamma _ {1 \gamma} \over V} = 2 \int \f {d^3 \, k} {( 2 \pi ) ^ 3} \, \f { \bar { \kappa } ( \vec {k} ) } 
{ e ^ { \omega / T  } - 1 } = \sum _ {p = 2} ^ {\infty} \, C _ p \, \f {2} {( 2 \pi ) ^ 2} \, T ^ {2 n} \, \Gamma (2 p) \, 
\f {\sqrt { \pi } \Gamma (p)} { \Gamma (p + 1 / 2) } \, \zeta(2p)  ,
\label{g1_d} 
\ee
where the factor of two accounts for two polarizations of the photon.

The expression in Eq.(\ref{g1_d}) can now be compared with the one resulting from the equations (\ref{g1_e}) and (\ref{dels}) thus determining the coefficients  $C _ p$ and yielding the expansion for $\bar \kappa$ in the form
\bea
\bar { \kappa } ( \vec {k} ) & = & \f {\a \, m^2} {4 \, \omega} \, \exp \l ( { - \f {\pi \, m ^ 2 } {e E} } \r )\, 
\sum _ {n = 1} ^ {\infty} \f {2 ^ {2 n + 4}} { \pi } \, \f {\Gamma ^ 2 ( n + 3 / 2 )} 
{ (2 n + 1) \, (n - 1)! \, (n+1)! \, (2 n + 1)!} \, \gamma _ \t ^ { 2 n }   \nonumber \\
& = & \f {\a \, m^2} {\omega} \, \exp \l ( { - \f {\pi \, m ^ 2 } {e E} } \r ) \, \sum _ {n = 1} ^ {\infty} \,  \f {\Gamma  ( n + 1 / 2 )} 
{ \sqrt{\pi} \, (n - 1)! \, n! \, (n+1)!} \, \gamma _ \t ^ { 2 n  }   \nonumber \\
& = & \f {\a \, m^2} {4 \, \omega} \,  \exp \l ( { - \f {\pi \, m ^ 2 } {e E} } \r ) \, \l ( \gamma _ \t ^ 2 + \f {1} {4} \gamma _ \t ^ 4 + \f {5} {192} \gamma _ \t ^ 6 + \dots \r ) 
\nonumber \\
& = & \f {\a \, m^2} {\omega} \, \exp \l ( { - \f {\pi \, m ^ 2 } {e E} } \r ) \,
\l [ I _ 1 (\gamma _ \t)  \r ] ^2 
\label{kbr}
\eea
with $\gamma_\theta$ being the Keldysh parameter $m \, \omega \, \sin \theta/(eE)$. The latter form of the result in Eq.(\ref{kbr}) in terms of the square of the Bessel function can be verified by squaring the standard Taylor expansion and collecting terms with the same power of the argument:
\bea
 \l [I_1(x) \r ]^2 & = & \l [ \sum_{k=0}^\infty {(x/2)^{2k+1} \over k! \, (k+1)!} \r ]^2 = \sum_{n=1}^\infty \, x^{2 n} \, \l [ 2^{-2n} \, \sum_{p=0}^{n-1} \, {1 \over p! \, (p+1)!\, (n-p-1)! \, (n-p)!} \r ] \nonumber \\
& = & \sum_{n=1}^\infty \, x^{2 n} \, {\Gamma(n+1/2) \over \sqrt{\pi} \, (n - 1)! \, n! \, (n+1)!}~.
\label{bess2}
\eea
The coefficients in the latter expansion clearly coincide with those in the second line of Eq.(\ref{kbr}).

\section{Discussion and conclusions.}
One can readily notice that the thermal expression (\ref{dels}) is applicable only at a low temperature $T < 1/(2R)$. However the resulting formula (\ref{kbr}) for the one-photon rate is valid at arbitrary values of the Keldysh parameter $\gamma_\theta$ (as long as the assumed bounds, $\omega \ll m$ and $eE \ll m^2$ are satisfied). This behavior, where the thermal expression is singular at the critical temperature, while the rates for individual processes are smooth functions, is similar to the one observed in analogous calculations in Refs.~\cite{mvsp,mvsw}.

It can be also readily argued that the absorption rate in Eq.(\ref{kbr}) is in fact related only to the photon polarization parallel to the electric field, so that for the photons with that particular polarization the absorption rate is twice larger than the average:
\be
\kappa_\parallel = 2 \, {\bar \kappa}~,
\ee
while for the photons with polarization orthogonal to the external field there is no absorption: $\kappa_\perp=0$.
Indeed, our Euclidean space calculation would not be affected if we considered the system, including the external electric field $E=E_x$, in a flat capacitor with small distance $\Delta$ between the plates (but still $\Delta \gg R$), i.e. if we imposed zero boundary condition on the components $A_y$ and   $A_z$ of the vector potential at $x = \pm \Delta/2$: $A_{y,z} (x=\pm \Delta/2)=0$. Clearly, such an arrangement leaves the components $A_x$ and $A_t$ intact, so that the potential created by the loop currents in the $(x,t)$ plane is still given by our Eq.(\ref{pota}), and one would arrive at the same result for the action per period $\Delta S$. On the other hand the boundary conditions at the plates of the capacitor introduce an energy gap $\pi/\Delta$ in the spectrum of the photons with polarization in the $y$ and $z$ direction and thus their presence in the thermal bath is suppressed. The absence of dependence of the thermal rate on the boundary conditions for the transversal to the external field $E$ polarizations implies that no absorption rate $\kappa_\perp$ arises for the perpendicular polarization as long as only the expansion in the leading parameter $\gamma_\theta$ is concerned. Such absorption would however arise in the next order of expansion in the ratio $\omega/m$. Within the described here technique the terms of that order originate from  the effects that are left beyond our essentially classical treatment of the electron Euclidean trajectory and of the field that it creates. In particular, the terms of higher order in $\omega/m$ would arise if one also includes the magnetic interaction of the current loops due to the spin of the electron.

Using the asymptotic expression for the Bessel function
\be
I _ 1 (z) = \f {e ^ z} { \sqrt { 2 \pi z } } , \mbox{ for } z \gg 1, 
\ee
one can find the exponential behavior of the probability rate (\ref{kbr})
\be
\bar {\kappa} \sim \exp \l [ - \f{m ^ 2} {eE} \l ( \pi - \f {2 \omega} {m} \sin \t \r ) \r ],
\ee 
which agrees with the $\omega \ll m$ limit of the exponential expression recently found in Ref.~\cite{Dunne:2009gi}.

In summary. We have calculated the rate of the photon-induced Schwinger process in the limit $\omega \ll m$ and $eE \ll m^2$ for arbitrary value of the Keldysh parameter $\gamma_\theta$. Our calculation differs from the one~\cite{Dunne:2009gi} based on the vacuum polarization operator in external field in that we use an extension of a  semiclassical treatment of the process to finite temperatures. The thermal rate is calculated in a standard way by considering the tunneling trajectory on a cylinder, i.e. periodic in the Euclidean time. The leading contribution of the photons, present in the thermal bath then arises from the the classical self-interaction of the electron current on the tunneling trajectory with itself on the cylinder. The contribution of stimulation of the pair creation by one photon is then determined from the term with appropriate power of the coupling $e^2$ in the thermal expression.  In this way we reproduce the nontrivial behavior in Eq.(\ref{kba}) of the calculated rate. We believe that the considered method, which we also recently applied in similar problems~\cite{mvsp,mvsw}, is of interest and can be used in other applications of tunneling processes induced by quantum particles.

\section*{Acknowledgments}
We thank Gerald Dunne, who brought to our attention the problem of the photon-stimulated Schwinger process, for illuminating discussions and for communicating to us his and his co-authors' result found by the standard technique.

The work of
A.M. is supported in part by the Stanwood Johnston grant from the Graduate
School of the University of Minnesota, RFBR Grant No. 07-02-00878 and by the
Scientific School Grant No. NSh-3036.2008.2.
The work of M.B.V. is supported in part by the DOE grant DE-FG02-94ER40823.

\end{document}